\documentclass{article}
\usepackage{graphicx}
\usepackage{hyperref}
\usepackage{psfrag}
\title{GRANULAR PRESSURE AND THE THICKNESS OF A LAYER JAMMING ON A ROUGH INCLINE}
\author{
Christophe Josserand,
 Pierre-Yves Lagr\'{e}e \and Daniel Lhuillier \\
 Laboratoire de Mod\'{e}lisation en M\'{e}canique, 
UMR CNRS 7607,\\ Bo\^ \i te 162,
Universit\'{e} Pierre et Marie Curie, 75252 Paris  FRANCE
}

\begin{document}

\maketitle

\begin{abstract}
Dense granular media have a compaction between the random loose and random close packings. For these dense media  
the concept of a granular pressure depending on compaction is not unanimously accepted 
because they are often in a "frozen" state which prevents them to explore all their 
possible microstates, 
a necessary condition for defining a pressure and a compressibility unambiguously. While periodic tapping or cyclic
fluidization
have already being used for that exploration, we here suggest that a succession of flowing states with velocities
slowly decreasing down
 to zero can also be used for that purpose. And we propose to deduce the pressure in \emph{dense and flowing} granular
 media from experiments
measuring the thickness of the granular layer that remains on a rough incline just after the flow has stopped.
\end{abstract}

\section{Introduction}
The existence of a pressure in granular media at rest is the simplest way to represent their stiffness.
 When the concentration of grains 
is above the random close packing, the granular medium acts as a poro-elastic solid and its pressure
 is a function not only of the compaction but also of the elastic constants of the material the grains are made of.
When the granular medium has a smaller compaction, in the range between the random loose and random close packings,
the grains can be considered as rigid and 
the expression of the granular pressure is far less evident. The difficulty comes from the glassy behaviour which makes
 it quite usual to find dense granular media in a \emph{frozen} state concerning their compaction. Hence the feeling 
 that the granular pressure is largely dependent on the way the medium was prepared. However, experiments
 have been conducted which aim at allowing the granular medium to reach a steady and quasi-equilibrium
 state concerning its compaction. 
These experiments relied on regular tappings or cyclic fluidization favouring the exploration
 of many microstates (see e.g. \cite{Now98} and \cite{Swi04}). 
We propose that a systematic exploration of the microstates can also be achieved
starting from a flowing granular medium, and \emph{slowly} reducing its velocity down to rest.
As a consequence, we suggest that some of the experiments on rough plates (inclined with angle $\theta$)
 which led to define the thickness $h_{stop}(\theta)$ which remains after the flow has stopped, 
can also be used to infer the relation 
between granular pressure and compaction.\\ While the pressure in poro-elastic media originates from elastic forces
 and granular deformations,
the pressure in dense and flowing granular media (those with a compaction between the random loose and 
random close packings) 
results mainly from the many possible random spatial configurations
of the grains. 
We will present the main features of this "randomness" pressure as well as those of  
 the more classical elastic pressure.
 Then some general expressions for the granular pression will be proposed. Finally two particular expressions will be deduced from experimental data concerning $h_{stop}$.
 
\section{The Configuration Pressure and the Disorder Pressure}
Consider a large volume containing many \emph{rigid} spheres with a high enough volume fraction
$\tilde \phi$ for a contact network to invade the whole volume.
 Let $\Omega (\tilde \phi) d \tilde \phi$ be the number of different spatial
 configurations of these spheres in the range between $\tilde \phi $ and
 $\tilde \phi + d\tilde \phi $. 
 To belong to $\Omega (\tilde \phi)$, a configuration must display a large enough number of contacts, but with zero
 forces at the contact points. The density of micro-states $\Omega (\tilde \phi)$ is thus a purely geometric concept. 
If instead of rigid spheres we were considering soft ones, we would say 
we are counting the number of configurations with zero energy ("incipient" contacts),
 yet able to resist an infinitesimal external pressure load. This density of states presumably vanishes
 below a minimum compaction $\phi_{min} \simeq 0.40$ (the gel threshold)
and above the maximum compaction
 $\phi_{max} \simeq 0.74$ (the most compact crystalline configuration). At some intermediate 
volume fraction $\phi _m$ the density of microstates displays a large maximum value. This 
intermediate compaction with the maximum number of microstates happens to be the loosest random packing.
According to Onoda and Liniger \cite{Ono98} its value for spherical grains is $\phi _m = 0.555$.
 Introduce now a number $P$ which represents a \emph{non-dimensional
 measure of the configuration pressure} and define the partition function
 \begin{eqnarray}
 Z(P) = \int _{\phi _{min}}^{\phi_{max}} e^{P \tilde \phi }\; \Omega (\tilde \phi) \; d \tilde \phi .\nonumber
 \end{eqnarray}
 The mean volume fraction is then related to the configuration pressure in the form:
$ \phi = \frac{d}{dP} \; Log \; Z .$
 This relation can be used to obtain  both $P(\phi)$ and
 the variance of the density fluctuations
$<(\tilde \phi - \phi )^2 > = d \phi / dP$. Another useful quantity is the configuration entropy
$S = \; Log \; Z - \; P\phi $ . This configuration entropy is a function of the mean volume fraction
and the non-dimensional configuration pressure is nothing but $P \;=\; - {dS}/{d\phi }.$
Many works starting from Kanatani \cite{Kan80} and revived by Edwards and Oakeshott \cite{Edw89}
 strived to find an explicit form for the
 partition function or for the configuration entropy. 
The general trends of their results are the followings : for a vanishing pressure, the mean volume fraction is
 $\phi_m$  while   for an infinite pressure, the volume fraction is $\phi_{max}$ 
and the compressibility $d\phi /dP$ vanishes.
 A very simple expression meeting these conditions is
\begin{eqnarray}
p^{configuration} \; \sim \; P \; \sim \;Log \frac{\phi_{max} - \phi _m}{\phi_{max} -\phi} .\nonumber
\end{eqnarray}
Close to the maximum volume fraction the configuration pressure has an expression quite similar 
to the pressure deduced from the entropy
 of the lattice-gas model (see e.g.
\cite{Wit04}). Note that the configuration pressure stems from the total number of different 
configurations, including both random and 
crystalline ones. It is also possible to select random configurations only. In this case one introduces the
 random close compaction $\phi _M$ above which all configurations display some cristalline order. For spheres, it is 
generally admitted that $\phi _M = 0.635$  and a possible expression for the disorder
 pressure is
 \begin{equation}
p^{disorder} \; \sim \;Log \frac{\phi_{M} - \phi _m}{\phi_{M} -\phi} .\label{eq:pdis}
\end{equation}
The gradient of the disorder pressure acts as a diffusion force which pushes the grains towards lower compactions,
those with a larger number of microstates. For the grains to have a chance to explore all microstates with
equal probability, the best solution is a steady flow. This is why the concept of a disorder granular pressure
 is more pertinent for dynamic situations and that it must be handled with care in static ones.
 The disorder pressure confers the granular medium a compressibility (variance of the compaction
fluctuations) which decreases with the compaction. When taking the above expression for granted the compressibility is 
proportional to $\phi _M - \phi $, 
in agreement with the experimental results of Nowak et al. \cite{Now98} but \emph{not} with those of Schroter et al.
\cite{Swi04} which display a minimum of the compressibility for a compaction between $\phi _m$ and $\phi _M$.
 The main features
 of $p^{disorder}$ are drawn schematically in Fig.1. 
\section{The Elastic Pressure}
The configuration or disorder pressure is a purely geometric (or entropic) concept. 
What matters is a minimum coordination
number between particles but the forces at the contact points are of no concern. It is then obvious that
 the configuration pressure is not the whole story. There is a second (and more intuitive) 
source of granular pressure which witnesses
 to the elastic stiffness of the granular medium and for which the magnitude of the contact forces is
 of utmost importance. That second
 contribution is the elastic   pressure $p^{elastic}$.
According to Aharonov and Sparks \cite{Aha99} and to O'Hern, Silbert, Liu and Nagel \cite{Her03} (see also 
Head and Doi \cite{Hea05} for a first principles approach) $p^{elastic}$ vanishes below a volume fraction $\phi_c$ which depends on the friction between grains and is \emph {sligthly smaller} than the random close packing $\phi_M$. Above $\phi_c$ it was found that 
 $
 p^{elastic}\sim (\phi - \phi_c)^{\alpha }
$  
with $\alpha =1$ for Hookean contact forces and
$\alpha = 3/2$ for Hertzian ones.
What is not apparent in the above expression is that
 the bulk modulus $dp^{elastic}/d\phi $ is
usually \emph {orders of magnitude larger} than $dp^{disorder}/d\phi $. When 
plotted on the same graph as $p^{disorder}$, it is as if $p^{elastic}$ was zero for $\phi <\phi _c$ 
and infinite for $\phi >\phi _c$, as
sketched in Fig.1 where the difference between $\phi _c$ and $\phi _M$ was exagerated for clarity. 
\section{The Rate-Independent Granular Pressure}
The pressure in a moving granular medium is generally made of two parts : a rate-dependent part 
which represents dilatancy effects in shear flows and a rate-independent part (hereafter noted $p$) representing 
the physical phenomena involved in both 
 $p^{elastic}$ and $p^{disorder}$. We are interested in the latter rate-independent contribution $p$.
 It seems clear that the very high stiffness for
$\phi \simeq  \phi_c$ is a feature of $p^{elastic}$ that must be shared by the total granular pressure. 
Much less obvious is
 the behaviour of $p$ for volume fractions slightly above $\phi_m$. Should we trust the linear behaviour displayed by  
the above expression for $p^{disorder}$ ? In fact, since our final aim is a comparison with experimental results,
it seems wise to introduce a general expression like
$p \sim (\phi - \phi _m)^n/(\phi _M -\phi)^m $
where the positive exponents $m$ and $n$ are left undetermined and we neglected the small difference between $\phi _c$ and
 $\phi _M$. Note that this expression for the rate-independent granular pressure 
holds in the very small range $\phi _m < \phi  < \phi _M$ specific of dense granular
media, and that the granular pressure vanishes for $\phi =\phi _m$ and diverges for $\phi =\phi _M$. In what follows
 we scale that
granular pressure with $\rho_p g D$ where $\rho_p$ is the mass per unit volume of the grains and $g$ is the 
acceleration of gravity. That scaling seems obvious when gravity is responsible for the confinment of the grains
but we checked \cite{Jos05} that the same scaling is also convenient when the confinment is due 
to an external pressure applied on the granular material.  The physical meaning of that special scaling is clear :
 the loads exerted on a granular medium with compaction in the range $\phi _m < \phi  < \phi _M$ are \emph{much
 smaller} than those leading to poro-elastic media with $ \phi  > \phi _M$. The latter ones involve elastic deformations
 of the grains while grains with a smaller compaction can be considered as rigid. Consequently we can not accept any 
scaling involving elastic constants and we are left with $\rho_p g D$ as the only relevant order 
of magnitude for stresses. Defining
the \emph{relative compaction} $\varphi$ as
\begin{equation}
\varphi \; = \; \frac {\phi - \phi _m}{\phi _M -\phi _m} ,
\end{equation}
we will test the potentialities of two expressions
for the rate-independent granular pressure in the range $0 \leq \varphi \leq 1$ :
 a rather general one with exponents $m>0$ and $n>0$ 
\begin{equation}
p \; = \; P_0 \;\rho_p g D \; \frac{\varphi ^n}{(1 - \varphi )^m},\label{eq:p1}
\end{equation}
and a special one identical to $p^{disorder}$ as given in (\ref{eq:pdis}) 
\begin{equation}
p \; = \; P_0 \; \rho_p g D \; Log \frac{1}{(1 - \varphi )}.\label{eq:p2}
\end{equation} 
That latter expression was already adopted in previous works \cite{Sav98,Jos04} and it suggests
that the elastic pressure plays a negligible role except for $\phi \simeq \phi _M$. We now
 consider an experiment likely to discriminate between (\ref{eq:p1}) and (\ref{eq:p2}) 
and able to give an order of magnitude for $P_0$.
\section{The Maximum Thickness of a Granular Layer that Stops on a Rough Incline}
Consider a layer of granular material flowing down a rough inclined  plate. Upon gently reducing the 
inclination, the layer ultimately stops at some angle $\theta$, with a thickness $h$. 
Since the flow velocity was slowly reduced to zero, the granular layer had time to explore a lot (if not all)
 of the microstates
involved in $p^{disorder}$. And since the flow was slow, the rate-dependent part of the granular pressure was already 
negligible before the layer jams. It is thus likely that the peculiar jammed state which the layer arrives at
is described by the rate-independent pressure $p$ defined above. The mechanical equilibrium of the 
\emph{freshly} jammed layer
 is thus expressed by
\begin{eqnarray}
0 = - \frac {\partial p}{\partial z} + \phi \rho _p g cos\theta , \quad  \mbox{and} \quad 
\tan{\theta } \; = \;min\;[\mu (z)] , 
\nonumber
\end{eqnarray}
where the $z-$axis is orthogonal to the free-surface of the layer and points downwards while $\mu $ is the friction
 coefficient which is possibly non-uniform over the layer thickness. Substituting
expression (\ref{eq:p1}), one deduces that the reduced compaction profile $\varphi (z)$ increases from zero close
 to the free-surface up to values close to one at a distance of order $L$ with 
$
\frac {L}{D} = \frac{P_0}{\phi _M cos\theta} \; .
$ 
Close to the free-surface
that is to say for $0 \leq z \ll L$,
and  far from the free-surface ($L \ll z$) the profile behaves asymptotically like
$\varphi (z) = (\frac {\phi_m}{\phi_M} \frac {z}{L})^{\frac {1}{n}}$ 
and $
\varphi (z) = 1 - (\frac{L}{z})^{\frac {1}{m}}$. 
Concerning the particular pressure (\ref{eq:p2}),
 the whole profile is exponential-like and given by
\begin{equation}
\varphi (z) = 1 - \frac {1}{1 + \frac{\phi _m}{\phi _M}(e^{\frac{z}{L}} - 1)} \; .\label{eq:fi}
\end{equation}
A similar "Fermi-Dirac" profile was already observed in experiments \cite{Cle91} and was also 
obtained in simulations of a frustrated lattice-gas model \cite{Nic97}. \\
Knowing the compaction profile, let us now focus on the layer thickness $h$. Da Cruz \cite{DaC04}
 has deduced from numerical simulations
 a very important (and a bit counter-intuitive) result : the
macroscopic friction coefficient \emph{decreases} almost linearly with the compaction and can be written as
\begin{equation}
\mu = tan \theta _{max} - (tan \theta _{max} - tan \theta _{min})\; \varphi \; .
\end{equation}
Since the compaction increases with the distance from the free-surface, \emph{the smallest value of $\mu$
happens very close to the rough plate}, that is to say for $z \simeq h$. Just after jamming, we thus have 
$tan \theta _{max} - (tan \theta _{max} - tan \theta _{min})\; \varphi (h_{stop}) = \tan \theta $. 
Taking the compaction profiles (\ref{eq:fipl}) into account, one deduces that when $\theta $
 slightly exceeds $\theta _{min}$ the layer is very thick and 
\begin{equation}
\frac {h_{stop}}{D} = \frac{P_0}{\phi _M cos\theta _{min}}
\left(\frac {\tan\theta _{max} - \tan\theta _{min}}{\tan \theta - \tan\theta _{min}}\right)^m .
\end{equation}
while when $ \theta$ comes close to $\theta _{max}$ the layer is thin and 
 \begin{equation}
\frac {h_{stop}}{D} = \frac{P_0}{\phi _m cos\theta _{max}}
\left(\frac {\tan\theta _{max} - \tan \theta}{\tan\theta _{max} - \tan\theta _{min}}\right)^n .
\end{equation}
 For the special case (\ref{eq:p2}), one obtains 
a rather simple expression that holds in
the whole range $\theta _{min} \leq  \theta \leq \theta _{max}$
 \begin{equation}
\frac {h_{stop}}{D} = \frac{P_0}{\phi _M cos\theta }\;
Log \left[1 + \frac{\phi_M}{\phi_m}\frac {\tan\theta _{max} - \tan \theta}
{\tan \theta - \tan\theta _{min}}\right] \; .
\end{equation}
\section{Experimental results}
Systematic measurements of the layer thickness were initiated by Pouliquen \cite{Pou99} who distinguished between
 the thickness $h_{start}(\theta)$ for an initially static layer and $h_{stop}(\theta)$ for an
 initially flowing layer. The experimental results 
were fitted by two different expressions 
 \begin{equation}
 \frac{h_{start,stop}(\theta )}{D} = B \; Log \; [\frac {\tan \theta _2 - \tan \theta _1 }{\tan \theta - \tan \theta _1}] \quad \mbox{or} \quad 
 \frac{h_{start,stop}(\theta )}{D} = B \;\frac {\tan \theta _2 - \tan \theta }{\tan \theta - \tan \theta _1} \; .
 \label{eq:h12}
 \end{equation}
 where $D$ is the grain size while $B$, $\theta _1$ and $\theta _2$ are constants, different
 for $h_{stop}$ and $h_{start}$.
Assuming that $\theta _2 = \theta _{max}$ and $\theta _1 = \theta _{min}$, these data fittings suggest 
two possible expressions for the granular pressure : either expression (\ref{eq:p2}) or expression (\ref{eq:p1})
with $n=1$ and $m=1$. However, most of the experimental values obtained for
 $B$, $\theta _1$ and $\theta _2$ must be considered
as non representative of the \emph{bulk} granular pressure because they were found to depend on
 the roughness of the inclined plate. In fact, the main difficulty
 with the interpretation of experimental results is to estimate the effective friction coefficient
 associated with this roughness. Our theoretical predictions were based on one main assumption : 
the minimum of effective friction occurs somewhere close to
 the incline, at the limit between the bulk and a thin boundary layer strongly influenced by the plate roughness.
 This supposes the effective friction in the boundary 
layer to be \emph{much larger} than its value in the bulk. In other words,
  to deduce the bulk granular pressure from experiments on inclines, 
we are led to exclude those experiments performed with relatively smooth plates, and more generally those for which 
the curves $h_{start,stop}(\theta )$ are strongly modified upon changing the plate roughness. And concerning those
 with a high enough roughness,
 we must exclude some boundary layer of thickness $\delta$ 
 and consider $h-\delta$ as the relevant thickness for bulk behaviour.
 Accordingly, we were led to discard all the experiments performed with glass beads because the friction generated
by the beads glued on the incline is only slightly larger than the friction in the bulk. But we considered as 
significative the experiments with
 sand flowing on carpets of various roughnesses. And for these experiments with sand we discarded 
a boundary layer with thickness estimated to $\delta \simeq 4D$. Because of the scarcity of data, we could not fully 
discriminate between expression (\ref{eq:p2}) and expression (\ref{eq:p1}) with $m=1$ and $n=1$. 
We could fit the experimental results for sand on carpets
(see Fig.2) with
 \begin{eqnarray}
\mu = 0.73 -  0.16 \varphi \quad \mbox{and} \quad
p \simeq \; 5.3 \;\rho _p g D \; Log \frac{1}{1 - \varphi } \;,\label{eq:p2f}
\end{eqnarray} 
as well as with
 \begin{eqnarray}
\mu = 0.73 -  0.20 \varphi \quad  \mbox{and} \quad
p \simeq \; 5.2 \;\rho _p g D \; \frac{\varphi }{1 - \varphi } \;.\label{eq:p1f}
\end{eqnarray} 
\section{Conclusions}
The concept of a rate-independent granular pressure 
 was proposed long ago for poro-elastic media \cite{Pas84}. We suggest this concept can be extended to dense and flowing 
 granular media for which this compaction-dependent pressure is the consequence of two distinct physical 
phenomena, of entropic and 
mechanistic nature respectively. We proposed to deduce the bulk granular pressure from experiments
on rough inclines and particularly those giving $h_{stop} (\theta )$. It happens that most
 of the experimental results concerning this
 thickness depend more on the roughness of the incline than on the bulk behaviour of the granular medium.
 However, the scarce results representative of the bulk behaviour can be interpreted with
 a granular pressure and an effective friction depending both on the compaction. Concerning sand, 
a rather satisfactory fit with the
 experimental results could be obtained with either expression (\ref{eq:p2f}) or expression (\ref{eq:p1f}), witnessing
 to the
 predominance of entropic effects in almost the whole compaction range $\phi _m <\phi <\phi _M$, while elastic effects
  come into play when approaching
  the random close packing $\phi _M$ only. \\
One may wonder why numerical simulations of flows down rough inclines (see e.g. \cite{Sil01}) predict flat compactions
profiles which are closer to those resulting from $p^{elastic}$ than $p^{disorder}$. 
 In fact, numerical simulations have so far described
 with great care the deterministic contact forces (hence the elastic stress) while they completely discarded the
stochastic processes associated with the exploration of the many microstates with contacts but zero contact forces. 
How to introduce these stochastic processes with the minimum change for the equations of motion ? 
A possibility is to write the equation of motion of particle $\alpha$ in the form
\begin{eqnarray}
m^{\alpha } \frac{d^2 \vec{R}^{\alpha }}{d t^2} \;=\; \sum_{\beta \neq \alpha}\vec{F}^{\alpha \beta} + \vec{F}^{\alpha } +
m^{\alpha } \vec{g}
\end{eqnarray}
where $\vec{F}^{\alpha \beta }$ is the force exerted at contact with particule $\beta$ while the stochastic force $\vec{F}^{\alpha }$ plays the role of the brownian force for thermal systems and
 witnesses to the random exploration of microstates. When tapping granular matter
one creates a series of short-lasting stochastic forces, while in steady flows this force is steadily acting.
There is no theory of $\vec{F}^{\alpha }$ we are aware of but, much like the brownian diffusion force 
 is the coarse-grained consequence of the stochastic brownian force, the gradient of the disorder
 pressure acts as a diffusion force which is the coarse-grained consequence of the granular stochastic forces. 
Hence, the concept of a rate-independent granular pressure
 can be used without moderation in the dense regime, provided this use is restricted to steadily flowing media.\\

\begin{figure} 
\psfrag{h}{$h_{stop}/D$}
\psfrag{th}{$\theta$}
\psfrag{pm}[bc]{ $\phi_m$}
\psfrag{pM}[bc]{ $\phi_M$} 
\psfrag{pc}[bc]{ $\phi_c$} 
\psfrag{p}{ $\phi$}
\psfrag{P}{ granular pressure}
\includegraphics[width=7cm]{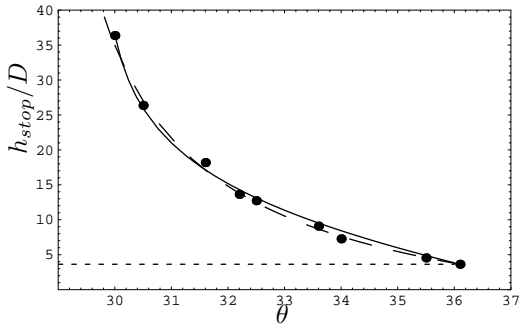} 
 \caption{ Schematic representation of the elastic pressure (dashed line) and disorder pressure (plain line) 
as a function of the compaction in the range between the random loose packing $\phi_m$ and the 
random close packing $\phi_M$. The disorder pressure vanishes below $\phi _m$ and diverges at $\phi _M$ 
while the elastic pressure vanishes
 below  $\phi_c$ and strongly increases above.
 }
\label{fig:1}
\end{figure}

\begin{figure} 
\psfrag{h}{$h_{stop}/D$}
\psfrag{th}{$\theta$}
\psfrag{pm}[bc]{ $\phi_m$}
\psfrag{pM}[bc]{ $\phi_M$} 
\psfrag{pc}[bc]{ $\phi_c$} 
\psfrag{p}{ $\phi$}
\psfrag{P}{ granular pressure}
\includegraphics[width=7cm]{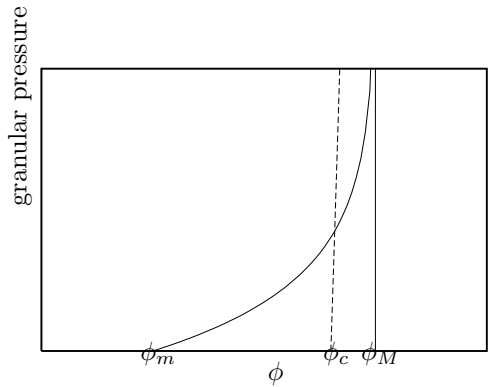} 
 \caption{ Dependence of $h_{stop}/D$ on the inclination $\theta $. Comparison between experimental results
 (points) for sand over carpets (from  \cite{DaC04}) 
and fitting curves deduced from the granular pressure (\ref{eq:p1f}) (plain) and (\ref{eq:p2f}) (dashed). 
Experimental results with 
$h_{stop}/D \leq  3.6$ were discarded and the value $\theta _{max}\approx 36.1$ was adopted.}
\label{fig:2}
\end{figure}

{\it Acknowledgments:} 
The authors appreciate interactions with R. Ball, F. Chevoir, D. Dean, 
J. Gollub, T. Harsley, 
D. Head, V. Kumaran, M. Nicodemi, J.N. Roux, G. Tarjus and P. Viot. D.L. thanks the Kavli Institute for Theoretical Physics for hospitality and support through the Granular
Physics Program GRANULAR05.

\end{document}